\documentclass[12pt,aps,prd,floats,floatfix,amsmath,nofootinbib,superscriptaddress]{revtex4}
\usepackage{setspace}
\usepackage{axodraw4j}
\usepackage{pstricks}
\usepackage{graphicx}
\usepackage{bm}
\usepackage{epsfig}
\usepackage{latexsym}
\usepackage{color}
\usepackage{colordvi}
\usepackage{verbatim}

\usepackage{axodraw4j}
\usepackage{pstricks}
\usepackage{color}

\newcommand{\beq}{\begin{equation}}
\newcommand{\eeq}{\end{equation}}
\newcommand{\bea}{\begin{eqnarray}}
\newcommand{\eea}{\end{eqnarray}}

\begin{document}

\title{Electroweak Corrections from Triplet Scalars}

\author{Zuhair U. Khandker, Daliang Li, and Witold Skiba}
\affiliation{Department of Physics, Yale University, New Haven, CT 06520}

\begin{abstract}
We compute the electroweak $S$ and $T$ parameters induced by $SU(2)_L$ triplet scalars up to one-loop order. 
We consider the most general renormalizable potential for a triplet and the Standard Model Higgs doublet. 
Our calculation is performed by integrating out the triplet at the one-loop level and also includes the one-loop renormalization group running. 
Effective field theory framework allows us to work in the phase with unbroken $SU(2)_L\times U(1)_Y$ symmetry.
Both $S$ and $T$ parameters exhibit decoupling when all dimensionful parameters are large while keeping dimensionless ratios fixed.
We use bounds on $S$ and $T$ to constrain the triplet mass and couplings.
\end{abstract}

\maketitle

\section{Introduction} \label{INTRO}

While the Standard Model (SM) Higgs boson has not been observed at the Large Hadron Collider with sufficient statistical significance, the allowed range of
Higgs masses is rapidly shrinking and there are preliminary hints of a Higgs boson with mass around 125~GeV~\cite{LHC}. The discovery of the Higgs boson
will certainly provide indirect information about  extensions of the SM\@. Precision electroweak corrections favor a light Higgs, so a heavier Higgs would indicate
new contributions to the Peskin-Takeuchi $S$ and $T$ parameters~\cite{PeskinTakeuchi}, and a 125~GeV Higgs means such contributions must be small. 

Here, we examine contributions to the $S$ and $T$ parameters arising from scalars transforming in the triplet representation of the $SU(2)_L$. 
Triplet scalars are a common ingredient of many extensions of the SM, such as GUTs, Little Higgs models, and seesaw models for neutrino masses.
In some instances, they also provide a cold dark matter candidate~\cite{CF05,CS07,FPb}. There are many other models that utilize triplet scalars. 

We consider heavy triplets with masses in the TeV range, or higher. We discuss two cases: triplets with either hypercharge 0 or 1. Such triplets can develop a vacuum 
expectation value without breaking the electromagnetic $U(1)$ and can have relevant couplings with the Higgs doublet. We consider the most general renormalizable potential for a triplet and the Higgs doublet.

The SM with a heavy triplet exhibits a hierarchy of scales characterized by the small parameter $\frac{v^2}{M^2}$, where $v$ is the electroweak scale and $M$ is the triplet mass.
This separation of scales motivates the use of an effective field theory (EFT) approach to study the triplet's effect on electroweak parameters. 
Accordingly, we integrate out the triplet at one-loop level and match to the SM with additional higher-dimensional operators $O_i$, with coefficients suppressed by appropriate powers of $M$.  The triplet's contribution to the $S$ and $T$ parameters is encoded in the coefficients of two higher-dimensional operators. 
We calculate these coefficients and also include their RG running from the matching scale, $M$, down to the electroweak scale, $v$. 
The logarithmic enhancement can be numerically relevant, although unlikely to be very important since large logarithms can only appear for very large triplet masses, in which case the triplet contributions to the $S$ and $T$ parameters are small anyway. Nevertheless, for completeness, we take RG evolution into account.

The triplet's contribution to $S$ and $T$ can be expanded in terms of $\frac{v^2}{M^2}$.  We work to leading order in this expansion, which allows for two important simplifications. First, only dimension 6 operators contribute at this order, and second, the masses of all SM fields can be set to zero. Since the $S$ and $T$ parameters are dimensionless, the higher-dimensional operators $O_i$ contribute to $S$ and $T$ proportionately to $\left(\frac{v}{M}\right)^{[O_i]-4}$, where $[O_i]$ is the dimension of $O_i$. Therefore, $\frac{v^2}{M^2}$ contributions come only from dimension 6 operators. Any contribution to dimension-6 operators from nonzero SM particle masses, which are proportional to the Higgs vev,  starts at order $\frac{1}{M^2}\frac{v^2}{M^2}$. Such a contribution would yield order $\frac{v^4}{M^4}$ terms for  $S$ and $T$, thus we neglect masses of SM fields.

Hence we perform all our calculations in the unbroken phase and avoid the complications of re-expressing fields in terms of mass eigenstates. This is an important difference from previous studies on the electroweak phenomenology of triplet scalars~\cite{RossVeltman, Gunion, Blank, LynnNardi, Czakon1, Czakon2, Forshaw, ForshawTalk, Chen03, Chen04, Chen05, Chen06, CP06, CW07, Chivukula, Chen08}. 
The EFT approach combined with the manifest $SU(2)_L\times U(1)_Y$ symmetry provides us with a transparent framework to systematically calculate all one-loop corrections to the $S$ and $T$ parameters coming from the triplet and obtain electroweak bounds on its mass and couplings. Our results can be used for analyzing electroweak constraints on SM extensions with a scalar triplet.

There are several articles in the literature, starting with \cite{Chen03, Chen05, Chen06} and later corroborated in \cite{CP06,Chivukula}, where it was argued that one-loop corrections to the $T$ parameter from triplet scalars do not decouple.  We find no such behavior. The results in \cite{Chen03, Chen04,Chen05, Chen06,CP06,CW07,Chivukula,Chen08} are obtained in the broken phase of the theory. In the EFT approach it is difficult to understand how a non-decoupling contribution may arise. There are no dimensionless parameters which grow with the triplet mass. There is a cubic scalar term, of mass dimension one, that is assumed to grow proportionately to the triplet mass, but ratios of mass parameters 
are assumed not to increase when the triplet mass increases.  Further discussion of  decoupling of triplets in an EFT language is contained in an Appendix of Ref.~\cite{S10}. 

This article is organized as follows. In the next section, we present the Lagrangian and sketch our approach. The main result of the paper are Eqs.~(\ref{FRSTN})-(\ref{FRSSC}) in Sec.~\ref{sec:results}.  Also in Sec.~\ref{sec:results}, we discuss electroweak constraints on the mass and couplings of the triplet. Details of the calculations are presented in three appendices.

\section{Methods}

\subsection{Lagrangian for a Triplet Scalar}

We consider the SM with an additional scalar field transforming as a triplet under the $SU(2)_L$. 
We restrict our attention to triplets with hypercharge of either $0$ or $\pm 1$, because with such choices the triplet can develop a vacuum expectation value (vev) without breaking the electromagnetic $U(1)$. This choice allows for relevant couplings between the triplet and the Higgs doublet.  At the electroweak scale, both the triplet and the Higgs doublet develop vevs. 
We integrate out the triplet fields above the electroweak scale obtaining higher-dimensional operators. As explained in the introduction, these operators are invariant under $SU(2)_L\times U(1)_Y$. The triplet dynamics, including its vev, are encoded in operators consisting of the Higgs and gauge fields.

We will refer to the real $0$-hypercharge triplet as  the neutral triplet and denote it by $\varphi^a$, and refer to the $(-1)$-hypercharge triplet as the charged triplet and denote it by $\phi^a$. The index $a$ is the $SU(2)_L$ index with $a=1,2,3$.  Since $\phi^{*a}$ has hypercharge $+1$ there is no reason to consider the $+1$ hypercharge fields separately. The covariant derivatives of these fields are 
\begin{eqnarray}
D_\mu \varphi^a &=& \partial_\mu \varphi^a + g_2 \epsilon^{abc} A_\mu^b \varphi^c, \\
D_\mu \phi^a &=& \partial_\mu \phi^a + g_2 \epsilon^{abc} A_\mu^b \phi^c + i g_1 B_\mu \phi^a, 
\end{eqnarray}
where $A_\mu^b$, $b=1,2,3$ and $B_\mu$ are the $SU(2)_L\times U(1)_Y$ gauge fields, while $g_2$ and  $g_1$ are the respective gauge couplings.

We consider gauge-invariant renormalizable couplings of the SM fields to either the neutral or to the  charged triplets
\begin{equation}
\mathcal{L}^{0} = \frac{1}{2}D_\mu \varphi^a D^\mu \varphi^a - \frac{M^2}{2} \varphi^a\varphi^a + \kappa H^\dagger \sigma^a H \varphi^a - \eta H^\dagger H \varphi^a \varphi^a + \mathcal{L}_{SM},
\label{L0}
\end{equation}  
\begin{eqnarray}
\mathcal{L}^{\pm1} &= &D_\mu \phi^{*a} D^\mu \phi^a - M^2 \left|\phi^a\right|^2 + \frac{\kappa}{2} \left(\tilde{H}^\dagger \sigma^a H\phi^a + h.c. \right) \nonumber \\
 && - \eta_1 H^\dagger H \phi^{*a} \phi^a - i\eta_2 H^\dagger \sigma^a H \epsilon_{abc} \phi^{*b}\phi^c + \mathcal{L}_{SM}.
\label{L1}
\end{eqnarray}
In the equations above, the superscripts on $\mathcal{L}$ denote the triplet hypercharge, $\sigma^a$'s are the Pauli matrices, $H$ is the Higgs doublet, $\tilde{H}=i\sigma_2 H^*$, and $\mathcal{L}_{SM}$ is the SM Lagrangian, whose Higgs and Yukawa sectors are given by
\begin{equation}
\mathcal{L}_{H+Yukawa} = D_\mu H^\dagger D^\mu H - \frac{\lambda}{4}\left(H^\dagger H\right)^2 - \left[y_T \, \bar{Q}_L \tilde{H} T_R + y_B\, \bar{Q}_L H B_R + h.c. \right].
\label{LH}
\end{equation}
$Q_{L}$ is the $SU(2)_L$ quark doublet consisting of the left-handed top and bottom fields, $T_R$ and $B_R$ are their right-handed counterparts, and $y_{T,B}$ are the Yukawa couplings. In Eq.~(\ref{LH}), we omit the light generations of quarks as well as the leptons since their Yukawa couplings 
are small. The only renormalizable coupling between triplets and SM fermions is a Yukawa coupling between a charged triplet and two left-handed lepton doublets. When the triplet gets a vev, such a term gives rise to a Majorana mass for the neutrino, hence the Yukawa coupling is small and, for our purposes, negligible. Finally, we omitted the possible triplet quartic couplings, ($\varphi_{a}\varphi_{a})^{2}$ in $\mathcal{L}_\varphi$, $\left(\phi^{a*}\phi^{a}\right)^{2}$  and $\phi^{a*}\phi^{a*}\phi^{b}\phi^{b}$ in $\mathcal{L}_\phi$, since these terms do not contribute to any electroweak observables at one loop. At the one-loop level, the quartics only contribute to the triplet mass renormalization, and these contributions are not observable. We simply assume that $M$ in Eqs.~(\ref{L0}) and (\ref{L1}) is the physical triplet mass.

Two of the terms in the Lagrangians above violate custodial symmetry, the cubic terms proportional to $\kappa$ in Eqs.~(\ref{L0}) and (\ref{L1}) and the quartic term proportional to $\eta_2$ in Eq.~(\ref{L1}), and therefore contribute to the $T$ parameter. The terms proportional to $\kappa$ contribute to $T$ starting at the tree level, while the term proportional to $\eta_2$ contributes to $T$ starting at the one-loop level. The $S$ parameter is generated at one loop and is generically small.


\subsection{EFT Approach to Calculating $S$ and $T$}  

Starting with the Lagrangian in Eq.~(\ref{L0}) or Eq.~(\ref{L1}), we integrate out the heavy triplet at the scale $\mu = M$ and match to an effective Lagrangian of the form
\begin{equation}
\mathcal{L}_{eff} = \mathcal{L}_{SM} + \sum_i a_i(\mu = M)\, O_i.
\label{Leff}
\end{equation}
Here, $\mathcal{L}_{SM}$ is the SM Lagragian and $\{O_{i}\}$ are $SU(3)\times SU(2)_L\times U(1)_Y$ gauge-invariant operators of dimension $6$ composed only of SM fields. At dimension 5, there is only one possible gauge-invariant operator---a term giving the left-handed neutrinos Majorana mass terms after electroweak symmetry breaking, which violates lepton number conservation and therefore must be very small. Moreover, this dimension-5 operator does not contribute to $S$ and $T$.  As we mentioned previously, since we only calculate the contribution to $S$ and $T$ to the leading order in $v^2/M^2$, we are only concerned with operators of dimension 6. All dimension 6 gauge-invariant and lepton- and baryon-number conserving operators that can appear on the RHS of Eq.~(\ref{Leff}) have been cataloged in~\cite{BW}. Of the 80 independent dimension-6 operators, we are interested in just two:
\begin{eqnarray}
O_S &=& H^\dagger \sigma^a H A_{\mu\nu}^a B^{\mu\nu}, \\
O_T &=& \left|H^\dagger D_\mu H\right|^2,
\end{eqnarray}
which are related to the $S$ and $T$ parameters. Letting $a_{S,T}$ denote the coefficients of $O_{S,T}$ in $\mathcal{L}_{eff}$, respectively, the measured values of the $S$ and $T$ parameters can be expressed in terms of these coefficients by~\cite{PeskinTakeuchi,HS04}
\begin{eqnarray}
S &= & \frac{4 v^2 \mathrm{sin}\,\theta_w \mathrm{cos}\,\theta_w}{\alpha}\, a_S(\mu=v)+\frac{1}{6\pi}\ln\frac{M_h}{M_{h,ref}}, \nonumber \\
T &= &-\frac{v^2}{2\alpha}\, a_T(\mu=v) -\frac{3}{8\pi\cos^{2}\theta_W}\ln\frac{M_h}{M_{h,ref}}, 
\label{ST}
\end{eqnarray}
where $v$ is the Higgs vev with $\left\langle H \right\rangle = \left( \begin{array}{c} 0\\ \frac{v}{\sqrt{2}} \end{array}\right)$, $M_h$ is the Higgs mass, $\theta_w$ is the weak mixing angle, $\alpha$ is the fine structure constant, and $v$ is the electroweak scale. The logarithmic terms encode the usual Higgs mass dependence of $S$ and $T$ in the SM.

We follow the standard EFT approach to obtain the low-energy values of the coefficients of effective operators. We integrate out the triplets at tree level and then at one loop and match to the effective
Lagrangian at the scale $\mu = M$. We then find the RG equations and evolve the couplings from $\mu=M$ down to $\mu = v$. More details of the calculations are presented in Appendices~\ref{app:matching} and \ref{app:running}, while an illustrative subset of the calculations is presented in Appendix~\ref{app:sample}.

\section{Results}
\label{sec:results}

\subsection*{Matching and Running}
Carrying out the procedure discussed in the previous section, we obtain the following coefficients $a_{S,T}$ of $O_{S,T}$ at the scale $v$:
\begin{align}
a_{T}^{0}(v) &= \frac{\kappa^{2}}{M^{4}}\left[-2 +  \frac{1}{(4\pi)^{2}}\left(-\frac{3}{2}\lambda+16\eta-\frac{37}{4}g_{2}^{2}+5\frac{\kappa^{2}}{M^{2}}\right)\right. \notag \\
& \left. -2\frac{1}{(4\pi)^{2}}\left(3\lambda-3g_{1}^{2}+\frac{9}{2}g_{2}^{2} + 24y_{B}^{2}+24y_{T}^{2}\right)\ln\left(\frac{v}{M}\right) \right],
\label{FRSTN}
\end{align}

\begin{align}
a_{T}^{\pm 1}(v) &= \frac{\kappa^{2}}{M^{4}}\left[1 + \frac{1}{(4\pi)^{2}}\left(\frac{3}{4}\lambda+\frac{11}{8}g_{1}^{2}+\frac{37}{8}g_{2}^{2}-\frac{17}{3}\frac{\kappa^{2}}{M^{2}}
 -\frac{22}{3}\eta_{2}-4\eta_{1}\right)\right. \notag \\
&  \left. +\frac{1}{(4\pi)^{2}}\left(3\lambda+\frac{3}{2}g_{1}^{2}+\frac{9}{2}g_{2}^{2} 
+24y_{B}^{2}+24y_{T}^{2}\right)\ln\left(\frac{v}{M}\right) \right]-\frac{2}{3} \frac{1}{(4\pi)^{2}}\frac{\eta_{2}^{2}}{M^{2}},
\label{FRSTC}
\end{align}

\begin{equation}
a_{S}^{0}(v)=\frac{1}{(4\pi)^{2}}\frac{g_{1}g_{2}}{M^{2}}\left[-\frac{1}{120}g_{2}^{2}-\frac{5}{24}\frac{\kappa^{2}}{M^{2}}
-\frac{1}{6}\frac{\kappa^{2}}{M^{2}}\ln\left(\frac{v}{M}\right)\right],
\label{FRSSN}
\end{equation}

\begin{equation}
a_{S}^{\pm 1}(v)=\frac{1}{(4\pi)^{2}}\frac{g_{1}g_{2}}{M^{2}}\left[\frac{1}{3}\eta_{2}-\frac{1}{40}g_{1}^{2}-\frac{1}{60}g_{2}^{2} 
+\frac{1}{8}\frac{\kappa^{2}}{M^{2}}+\frac{1}{3}\frac{\kappa^{2}}{M^{2}}\ln\left(\frac{v}
{M}\right)\right],
\label{FRSSC}
\end{equation}
where the superscripts on the coefficients $a_{S,T}$ indicate the triplet hypercharge. 

\subsection*{Exclusion Plots}

We now turn to the experimental bounds and illustrate the allowed regions of parameters for triplets. The results in Eqs.~(\ref{FRSTN})-(\ref{FRSSC}) are converted into the corresponding values of the $S$ and $T$ parameters according to Eq.~(\ref{ST}).  We use the $95\%$ confidence level limits on $S$ and $T$  obtained by the Gfitter group in Ref.~\cite{Baak:2011ze}, taking the top mass to be 173~GeV and the Higgs mass to be 125~GeV, to constrain  the masses and couplings of  the triplet scalars.

\begin{figure}[htb]
\includegraphics[width=13cm]{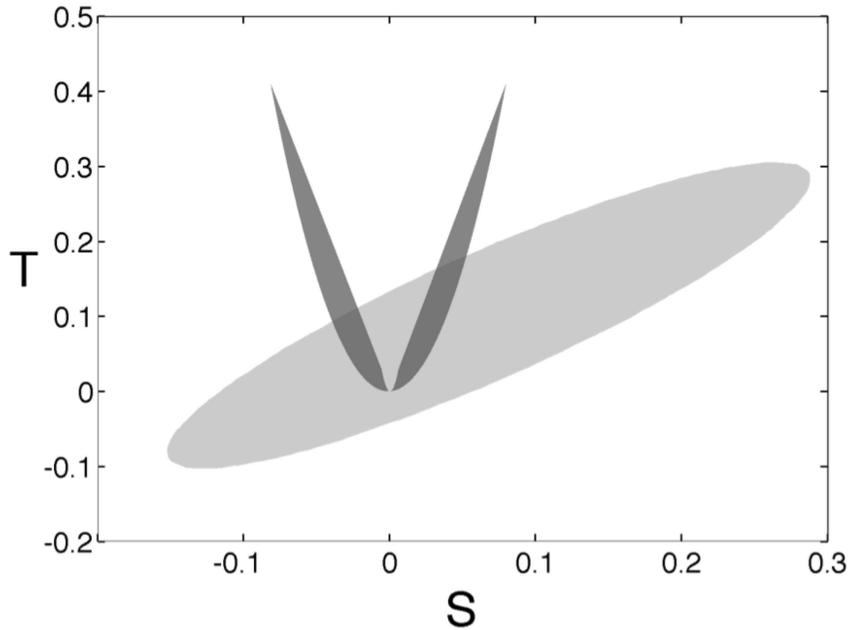}
\caption{The $\eta_2$-dependent contributions to $S$ and $T$ from the charged triplet. The dark gray region shows the triplet's contribution when the triplet mass M and coupling constant $\eta_{2}$ are scanned over the region $400~{\rm GeV}\le M\le1500~{\rm GeV}$ and $ -2\le\eta_{2}\le2$ after setting $\kappa=0$, $\eta_{1}=0$, and  $M_{h}=125~{\rm GeV}$. The light gray region illustrates the 95\% confidence region of allowed values for the $S$ and $T$ parameters~\cite{Baak:2011ze}. \label{CSC} }
\end{figure}

For both the neutral and the charged triplet, contributions to $T$ arise already at tree level while contributions to $S$ arise at loop level,  thus the $S$ parameter will generically be much smaller than the $T$ parameter. For the neutral triplet, the tree-level contribution to $T$ is positive. Such positive contributions can accommodate larger Higgs masses in the fit to electroweak data, for example if  the recent hints of the Higgs boson around 125~GeV~\cite{LHC} turn out to be false. 

\begin{figure}[htb]
\includegraphics[width=13cm]{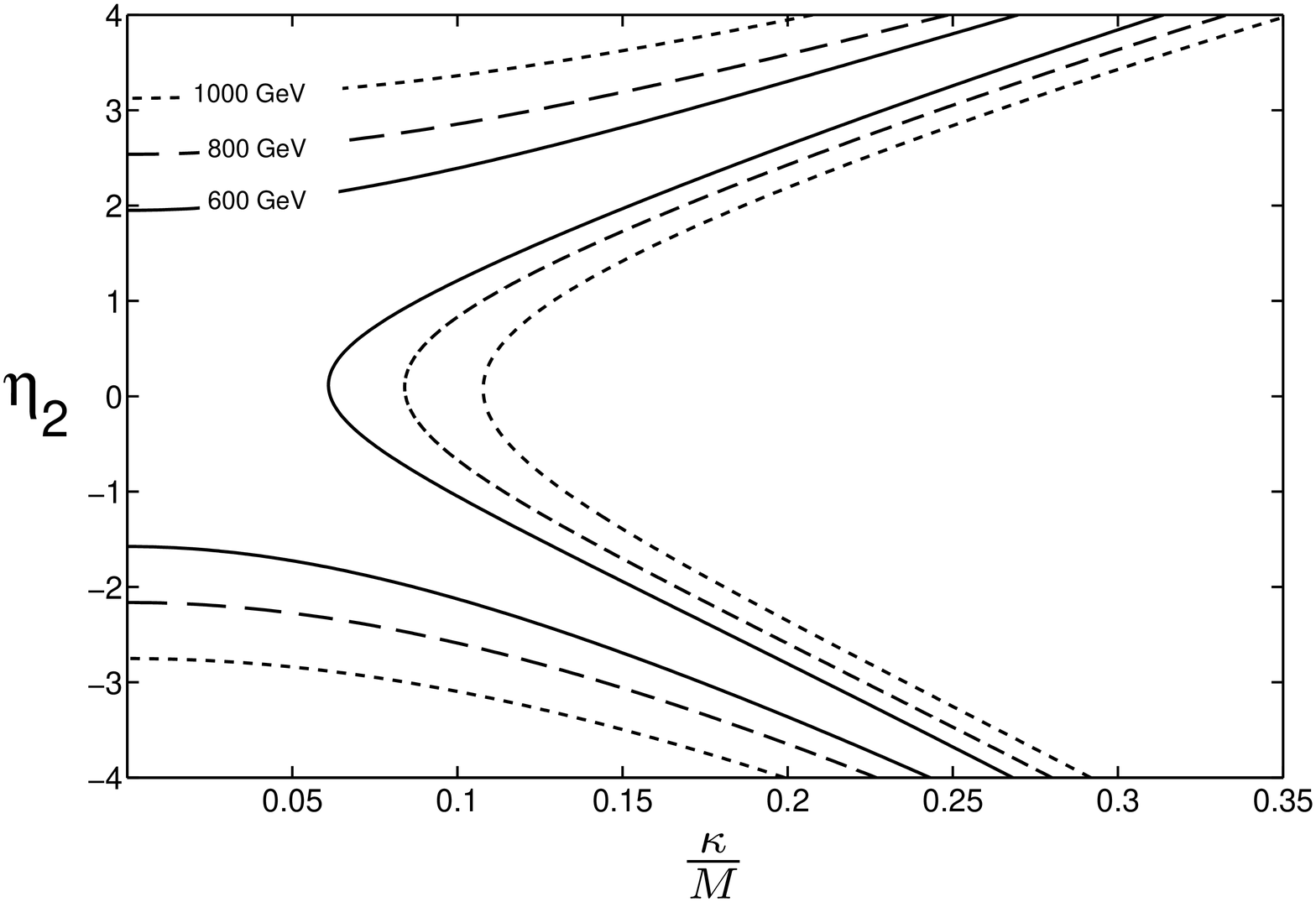}
\caption{The $95\%$ exclusion regions in the $\eta_{2}$-$\kappa$ plane for different masses of the charged triplet assuming $M_{h}=125~{\rm GeV}$. The allowed ranges lie between the curves corresponding to a given triplet mass. For certain ratios of $\eta_{2}$ to $\frac{\kappa}{M}$, relatively large values of these parameters are consistent  with experimental constraints. This is because these two contributions nearly cancel each other at such ratios. \label{CKE}}
\end{figure}

The charged scalar exhibits a new feature which is absent in the neutral case. In the charged-scalar Lagrangian, Eq.~(\ref{L1}), in addition to the cubic interaction proportional to $\kappa$, the interaction proportional to $\eta_2$ also violates custodial symmetry. An analogue of this $\eta_2$ interaction is absent in the neutral-scalar Lagrangian. The interaction term proportional to $\eta_2$ generates a one-loop contribution to $T$ that is positive, proportional to $\eta_2^2$, and independent of $\kappa$. Fig.~\ref{CSC} illustrates the $\eta_2^2$ contribution. 

\begin{figure}[htb]
\includegraphics[width=13cm]{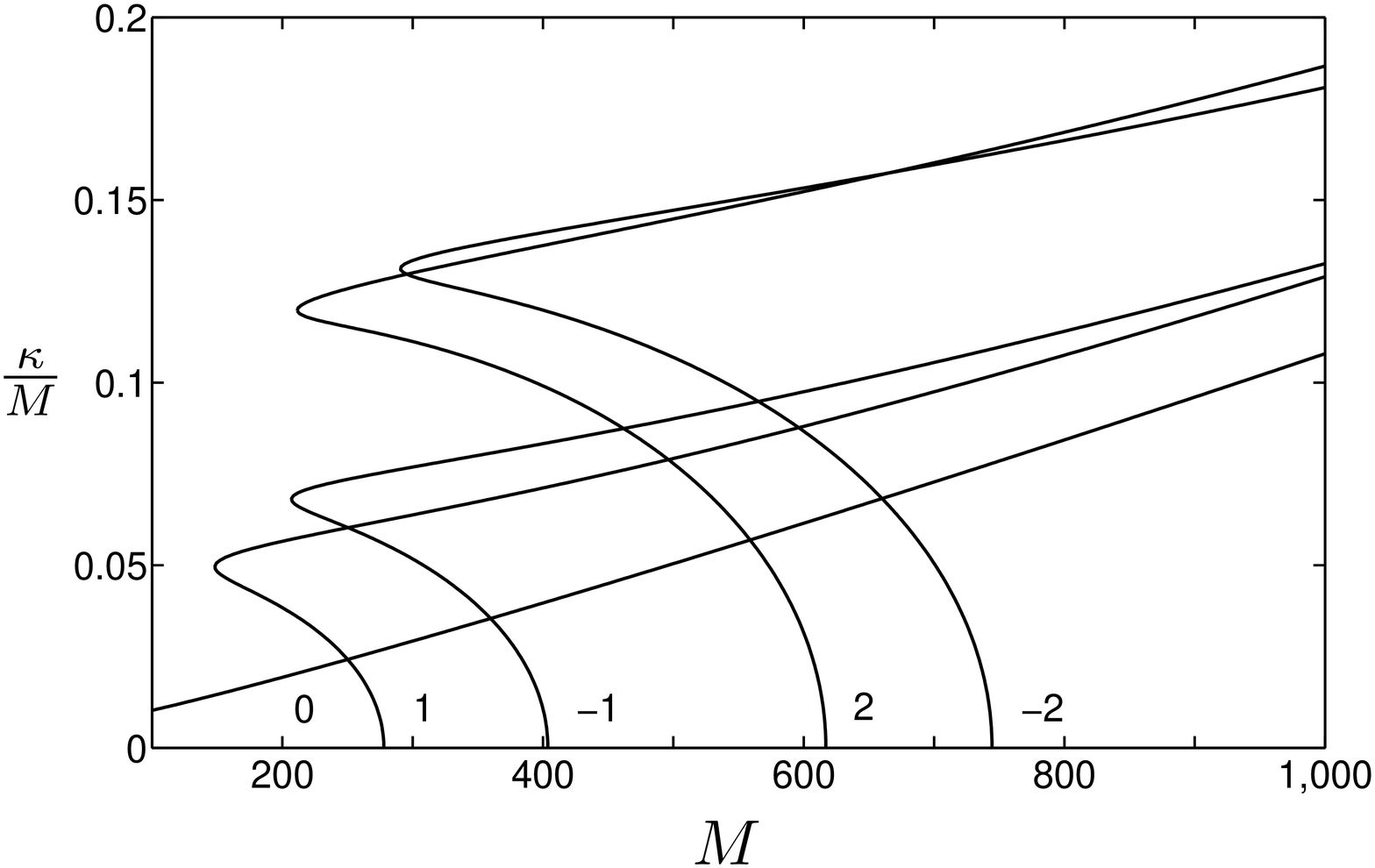}
\caption{The 95\% exclusion regions of $\frac{\kappa}{M}$ and $M$ for different values of $\eta_2$ in the charged triplet case. We set  $\eta_1=0$ and $M_{h}=125~{\rm GeV}$. The $\eta_2$ value for each curve is labelled at the bottom. The allowed regions are to the right of each curve.  \label{CKOMM}}
\end{figure}

\begin{figure}[hbt!]
\includegraphics[width=13cm]{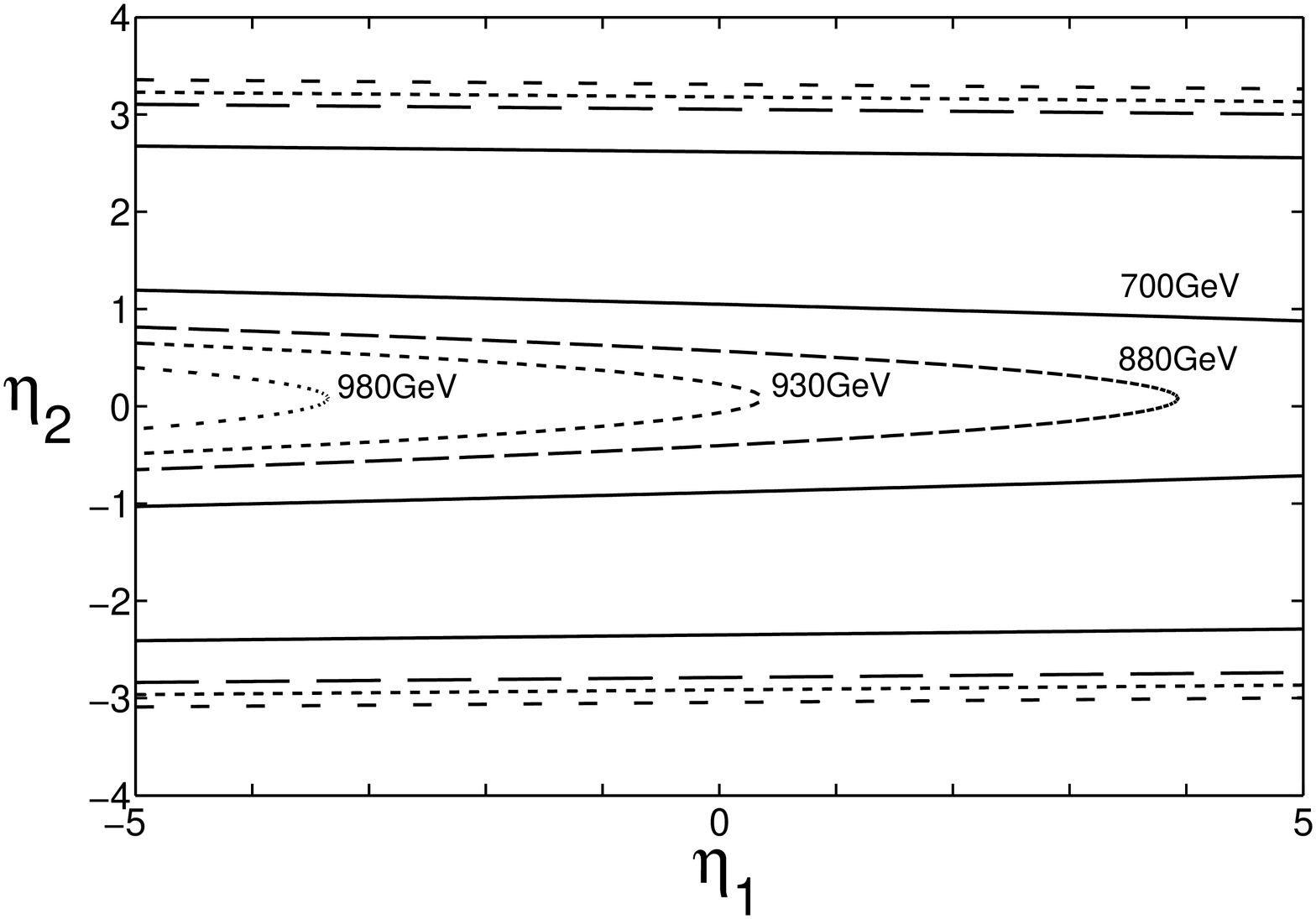}
\caption{The 95\% exclusion regions of $\eta_1$ and $\eta_2$ for different charged triplet masses, M, where we have fixed $\frac{\kappa}{M}=0.1$ and $M_h=125~{\rm GeV}$. The allowed region lies between the top- and bottom-most lines corresponding to a given triplet mass and to the right of the corresponding curve in the middle.  \label{CEE}}
\end{figure}

The presence of this $\eta_2^2$ contribution has a number of consequences. First, the positive 1-loop, $\eta_2^2$ contribution to $T$ can compete with the negative, tree-level $\kappa^2$ contribution, especially for small values of $\frac{\kappa}{M}$. This is shown in Fig.~\ref{CKE}. Second, the allowed $M$ versus $\frac{\kappa}{M}$ parameter space is modified by the $\eta_2^2$ contribution compared to the neutral case. The importance of the $\eta_2$ contribution is largest for small values of $M$ and $\frac{\kappa}{M}$, as illustrated in Fig.~\ref{CKOMM}. Finally, even for fixed but small $\frac{\kappa}{M}$, the $\eta_2^2$ contribution leads to a nontrivial $\eta_1$ versus $\eta_2$ allowed parameter space. This is shown in Fig.~\ref{CEE}.

\section{Conclusions}
\label{sec:conclusion}

We calculated the corrections to the $S$ and $T$ parameters induced by electroweak triplet scalars up to one-loop order. We considered the most general renormalizable Lagrangian for a triplet scalar coupled to the SM Higgs doublet. We computed the $S$ and $T$ parameters in an effective theory in which the triplets are integrated out 
by considering the corresponding operators of dimension 6, that is we worked to the leading order in $v^2/M^2$. Our results are contained in Eqs.~(\ref{FRSTN}) through (\ref{FRSSC}).

There are two reasons for performing this calculation. First, it is useful for constraining the parameter space of the triplets. In most cases, the tree-level contribution to $T$ dominates the corrections to the oblique parameters. This dominant correction is proportional to the cubic coupling of the triplet to Higgs doublets in Eqs.~(\ref{L0}) and (\ref{L1}). When the cubic coupling is small the loop effects can be significant. There are 1-loop contributions to $S$ that are independent of the cubic coupling, and for the charged triplet there is also a quartic coupling that contributes to $T$ independently of the cubic coupling. 

The second reason is that there are several results in the literature in Refs.~\cite{Chen03,Chen04,Chen05,Chen06}, \cite{CP06,CW07}, and \cite{Chivukula} that find that the corrections from the triplets do not decouple in the limit of large triplet masses at the one-loop level. If true, this is of important consequence for triplet phenomenology. However, we find no such behavior and the $S$ and $T$ parameters approach zero for large triplet mass. The cubic coupling, $\kappa$, between the triplet and the Higgs doublet involves a dimensionful  constant. As in the references above, we assume that the dimensionless ratio $\frac{\kappa}{M}$ does not increase with $M$,  that is in the large $M$ limit $\kappa$ does not grow faster than $M$. 

The calculations in Refs.~\cite{Chen03,Chen04,Chen05,Chen06,CP06,CW07,Chivukula} were performed in the broken phase, in which the triplet and the doublet acquire vevs. We work in the unbroken phase of the theory. It is not clear to us why these two approaches would give different answers. Decoupling is not at all surprising in the effective theory. 
The $S$ and $T$ parameters correspond to dimension-6 operators and are thus inversely proportional to the triplet mass squared. Dimensionful couplings, like $\kappa$, can only enter in the ratio $\frac{\kappa}{M}$, and cannot appear in ratios with a light scale, for example as $\frac{\kappa}{v}$. This statement is independent of the loop order.

One might  be leery of a result obtained in the unbroken phase. Of course, this should not be an issue as symmetry breaking is a low-energy effect. A properly constructed effective theory matches the infrared behavior of the full theory. A partial result for the $T$ parameter was presented in Ref.~\cite{S10}, where it was explicitly shown how the infrared divergencies match between the full and effective theories when the triplet is decoupled at one loop.   In other words, the coefficients of effective operators are independent of the Higgs vev and therefore can be computed assuming a vanishing vev. 

The unbroken phase calculation offers one advantage---it is less complicated. There is no need to find mass eigenstates and no need to re-express interactions in terms of mass eigenstates. This is obviously a computational issue that cannot be responsible for the discrepancy of the results. Some speculations as to why apparently non-decoupling behavior occurs in the broken-phase calculations were presented in Ref.~\cite{S10}. At the moment, we have no further insights into the underlying cause of the discrepancy. 

\section*{Acknowledgements}
We thank W. Goldberger for discussions.  This work is supported in part by DOE grant DE-FG-02-92ER40704.

\appendix
\section{Matching}
\label{app:matching}
In this appendix, we describe the procedure for integrating out the heavy triplet. In order to match the effective Lagrangian, Eq.~(\ref{Leff}), to the full Lagrangian, either in Eq.~(\ref{L0}) or in Eq.~(\ref{L1}), we need to determine the coefficients $a_i$ in the effective theory such that Green's functions in the full and effective theories are identical to the desired accuracy. For any scattering amplitude $G$ in which triplets do not appear in the external states, the matching condition is
\begin{equation}
G_{full} = G_{eff}\left( {a_i} \right),
\label{MC}
\end{equation}
where the subscripts $full$ and $eff$ denote the amplitudes calculated in either the full or the effective theory, respectively. 
Both sides of Eq.~(\ref{MC}) can be expanded in loop orders. Let $a_{i}=a_{i}^{tree}+a_{i}^{1-loop}+\ldots $, and similarly for $G$. Up to 1-loop order, the condition (\ref{MC}) becomes   
\begin{equation}
G_{full}^{tree} =  G_{eff}^{tree}({a_{i}^{tree})},
\label{MC0}
\end{equation}
\begin{equation}
G_{full}^{1-loop} =  G_{eff}^{tree}({a_{i}^{1-loop})}+G_{eff}^{1-loop}({a_{i}^{tree})}.
\label{MC1}
\end{equation}
In the following, we will use Eqs.~(\ref{MC0}) and (\ref{MC1}) to determine $a_{S,T}^{tree}$ and $a_{S, T}^{1-loop}$ at the matching scale $\mu=M$.


\subsection*{Tree Level}
Because the triplets have significant couplings only  to the gauge bosons and the Higgs  we are interested in oblique corrections in $\mathcal{L}_{eff}$, that is in operators without fermions. At tree-level, all full-theory topologies involving the triplet and either Higgs or gauge-boson external lines are shown in Fig.~\ref{tree}. Integrating out the triplet from these diagrams induces the following effective operators, up to dimension six:
\begin{eqnarray}
&& O_1 \equiv \frac{1}{2}\left(D^2 H^\dagger H H^\dagger H + h.c.\right), \hspace{5mm} O_2 \equiv D_\mu H^\dagger D^\mu H H^\dagger H, \hspace{5mm} \nonumber \\
&&O_T = \left|H^\dagger D_\mu H\right|^2, \hspace{5mm}   \left(H^\dagger H\right)^2, \hspace{5mm} \left(H^\dagger H\right)^3.
\end{eqnarray}

We can ignore $\left(H^\dagger H\right)^2$ and $\left(H^\dagger H\right)^3$. Contributions to $\left(H^\dagger H\right)^2$ simply renormalize an existing term in Eq.~(\ref{LH}), while the $\left(H^\dagger H\right)^3$ operator can be ignored because it contributes neither to the matching nor to the one-loop RG running of $a_{S,T}$. This leaves us with $O_i$, $i=1,2,T$, so that the effective Lagrangian takes the form
\begin{equation}
\mathcal{L}_{eff} = \mathcal{L}_{SM} + a_i O_i.
\label{app:Leff}
\end{equation}
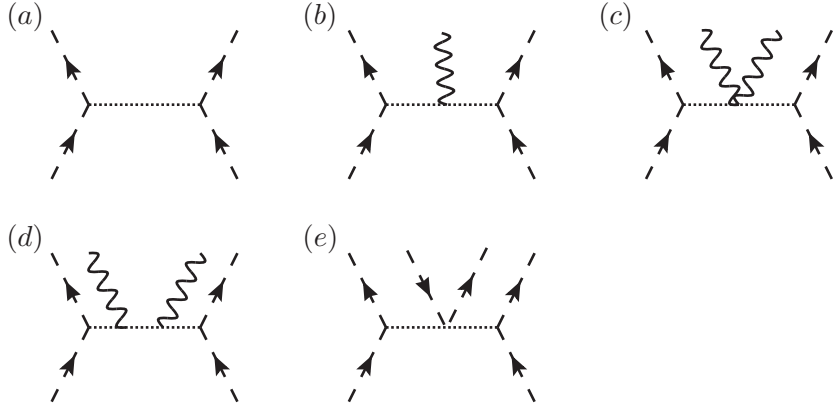
\begin{figure}
\fcolorbox{white}{white}{
  \begin{picture}(306,162) (31,-35)
    \SetWidth{1.0}
    \SetColor{Black}
    \Line[dash,dashsize=4,arrow,arrowpos=0.5,arrowlength=5,arrowwidth=2,arrowinset=0.2](42,50)(56,78)
    \Line[dash,dashsize=4,arrow,arrowpos=0.5,arrowlength=5,arrowwidth=2,arrowinset=0.2](56,78)(42,106)
    \Line[dash,dashsize=1](56,78)(98,78)
    \Line[dash,dashsize=4,arrow,arrowpos=0.5,arrowlength=5,arrowwidth=2,arrowinset=0.2](98,78)(112,106)
    \Line[dash,dashsize=4,arrow,arrowpos=0.5,arrowlength=5,arrowwidth=2,arrowinset=0.2](112,50)(98,78)
    \Line[dash,dashsize=4,arrow,arrowpos=0.5,arrowlength=5,arrowwidth=2,arrowinset=0.1](168,78)(154,106)
    \Line[dash,dashsize=4,arrow,arrowpos=0.5,arrowlength=5,arrowwidth=2,arrowinset=0.2](154,50)(168,78)
    \Line[dash,dashsize=1](168,78)(210,78)
    \Line[dash,dashsize=4,arrow,arrowpos=0.5,arrowlength=5,arrowwidth=2,arrowinset=0.2](210,78)(224,106)
    \Line[dash,dashsize=4,arrow,arrowpos=0.5,arrowlength=5,arrowwidth=2,arrowinset=0.2](224,50)(210,78)
    \Line[dash,dashsize=4,arrow,arrowpos=0.5,arrowlength=5,arrowwidth=2,arrowinset=0.1](280,78)(266,106)
    \Line[dash,dashsize=1](280,78)(322,78)
    \Line[dash,dashsize=4,arrow,arrowpos=0.5,arrowlength=5,arrowwidth=2,arrowinset=0.2](322,78)(336,106)
    \Line[dash,dashsize=4,arrow,arrowpos=0.5,arrowlength=5,arrowwidth=2,arrowinset=0.2](266,50)(280,78)
    \Line[dash,dashsize=4,arrow,arrowpos=0.5,arrowlength=5,arrowwidth=2,arrowinset=0.2](336,50)(322,78)
    \Line[dash,dashsize=4,arrow,arrowpos=0.5,arrowlength=5,arrowwidth=2,arrowinset=0.1](56,-6)(42,22)
    \Line[dash,dashsize=4,arrow,arrowpos=0.5,arrowlength=5,arrowwidth=2,arrowinset=0.2](42,-34)(56,-6)
    \Line[dash,dashsize=1](56,-6)(98,-6)
    \Line[dash,dashsize=4,arrow,arrowpos=0.5,arrowlength=5,arrowwidth=2,arrowinset=0.2](98,-6)(112,22)
    \Line[dash,dashsize=4,arrow,arrowpos=0.5,arrowlength=5,arrowwidth=2,arrowinset=0.2](112,-34)(98,-6)
    \Line[dash,dashsize=4,arrow,arrowpos=0.5,arrowlength=5,arrowwidth=2,arrowinset=0.1](168,-6)(154,22)
    \Line[dash,dashsize=4,arrow,arrowpos=0.5,arrowlength=5,arrowwidth=2,arrowinset=0.2](154,-34)(168,-6)
    \Line[dash,dashsize=1](168,-6)(210,-6)
    \Line[dash,dashsize=4,arrow,arrowpos=0.5,arrowlength=5,arrowwidth=2,arrowinset=0.2](210,-6)(224,22)
    \Line[dash,dashsize=4,arrow,arrowpos=0.5,arrowlength=5,arrowwidth=2,arrowinset=0.2](224,-34)(210,-6)
    \Line[dash,dashsize=4,arrow,arrowpos=0.5,arrowlength=5,arrowwidth=2,arrowinset=0.2](192,-4)(206,24)
    \Line[dash,dashsize=4,arrow,arrowpos=0.5,arrowlength=5,arrowwidth=2,arrowinset=0.1,flip](190,-5)(176,23)
    \Photon(189,105)(191,78){3}{4}
    \Photon(287,106)(301,78){3}{4}
    \Photon(301,78)(315,106){3}{4}
    \Photon(70,-6)(56,22){3}{4}
    \Photon(84,-6)(98,22){3}{4}
    \Text(25,106)[lb]{\small{\Black{$(a)$}}}
    \Text(137,106)[lb]{\small{\Black{$(b)$}}}
    \Text(249,106)[lb]{\small{\Black{$(c)$}}}
    \Text(25,22)[lb]{\small{\Black{$(d)$}}}
    \Text(137,22)[lb]{\small{\Black{$(e)$}}}
  \end{picture}
}
\caption{Tree-level diagrams of the full theory contributing to oblique operators in the effective theory (neutral triplet case). The longer dashed lines represent Higgs fields, while the shorter dashed lines represent the heavy triplet.}
\label{tree}
\end{figure}

To determine the matching coefficients, it suffices to consider only the diagram in Fig.~\ref{tree}(a). We can ignore Fig.~\ref{tree}(e), because it only contributes to the operator $\left(H^\dagger H\right)^3$ (and to other operators with dimensions larger than six). We can ignore Fig.~\ref{tree}(b)-(d), because they are related by gauge invariance to Fig.~\ref{tree}(a). For example, consider $O_1 = \frac{1}{2}\left(\partial^2 H^\dagger H H^\dagger H + h.c.\right) + \text{gauge interactions}$. The form of  vertices with gauge bosons is fixed by gauge invariance and follows from making the derivatives covariant. To match the full theory to $O_1$, it suffices to find the contribution to $\frac{1}{2}\left(\partial^2 H^\dagger H H^\dagger H + h.c.\right)$, for which only Fig.~\ref{tree}(a) is pertinent. (Conversely, one could  use Fig.~\ref{tree}(b)-(d) to match to the gauge interaction parts of $O_1$. This equivalent matching procedure is discussed in \cite{S10}.) This is possible because we take advantage of the full electroweak gauge symmetry.

\begin{figure}
\fcolorbox{white}{white}{
  \begin{picture}(289,136) (275,-101)
    \SetWidth{1.0}
    \SetColor{Black}
    \Line[dash,dashsize=4,arrow,arrowpos=0.5,arrowlength=5,arrowwidth=2,arrowinset=0.2](432,-88)(464,-56)
    \Line[dash,dashsize=4,arrow,arrowpos=0.5,arrowlength=5,arrowwidth=2,arrowinset=0.2](496,-56)(528,-88)
    \Line[dash,dashsize=4,arrow,arrowpos=0.5,arrowlength=5,arrowwidth=2,arrowinset=0.2](496,-24)(528,8)
    \Line[dash,dashsize=4,arrow,arrowpos=0.5,arrowlength=5,arrowwidth=2,arrowinset=0.2](432,8)(464,-24)
    \Text(272,-48)[lb]{\large{\Black{$G_{s_1 , s_2 , s_3 , s_4}(p_1 , p_2 ,p_3 ,p_4 )=$}}}
    \Text(415,14)[lb]{\small{\Black{$H_{s_1},p_1$}}}
    \Text(415,-106)[lb]{\small{\Black{$H_{s_ 2},p_2$}}}
    \Text(529,13)[lb]{\small{\Black{$H_{s_3},p_3$}}}
    \Text(529,-102)[lb]{\small{\Black{$H_{s_4},p_4$}}}
    \GOval(480,-41)(24,24)(0){0.882}
  \end{picture}
}
\caption{The extraction of the coefficient $a_T$ uses amplitude $G$ as defined in this figure.}
\label{2Higgs}
\end{figure}
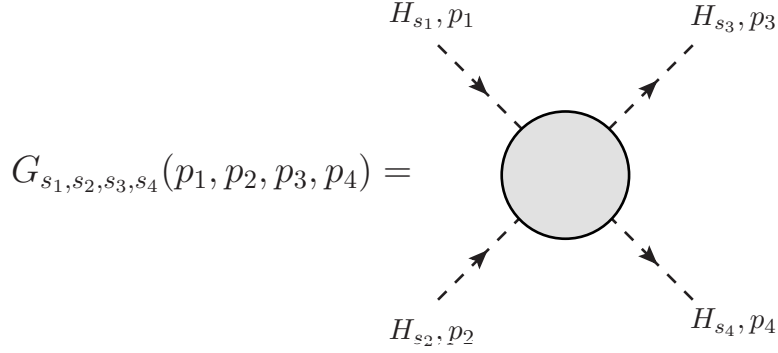

When the triplet is integrated out, all three $O_i$ in Eq.~(\ref{app:Leff}) receive nonzero contributions. We can determine the contribution to each operator using three different configurations of external momenta and components of the Higgs doublets on the external lines in Fig.~\ref{tree}(a). Specifically, as shown in Fig.~\ref{2Higgs}, we define $G_{s_1s_2s_3s_4}\left(p_1,p_2,p_3,p_4\right)$ to be the 2-Higgs to 2-Higgs scattering amplitude where the two incoming Higgs fields have momenta $\{p_1,p_2\}$ and components $\{s_1,s_2\}$, while the outgoing Higgses have $\left\{p_3,p_4\right\}$ and $\left\{s_3,s_4\right\}$. In our notation, $s_j=1$ means the upper component of the Higgs doublet on the $j$-th line, while $s_j=2$ means the lower component. Different operators $O_i$ have different dependence on the momenta and different contractions of the Higgs fields,
so choosing different configurations allows us to extract the coefficients of independent operators from the same diagram. 

We choose the three different configurations of $\{p_j,s_j\}$ to be: 
\begin{equation}
G_{1} \equiv G_{1212}\left(p,0,p,0\right), \hspace{5mm} G_{2} \equiv G_{1212}\left(p,0,0,p\right), \hspace{5mm} G_{3} \equiv G_{1212}\left(p,-p,0,0\right).
\label{G}
\end{equation}  
At the matching scale, the tree-level values of the EFT coefficients $a_i$ are determined by: 
\begin{alignat}{3}
G_{1,full}^{tree} - G_{3,full}^{tree}  \hspace{3mm} &= \hspace{3mm}G^{tree}_{1,eff}\left(a_i^{tree}\right) - G^{tree}_{3,eff}\left(a_i^{tree}\right) \hspace{3mm}&= \ldots+ i p^2 a_2^{tree}  + \ldots  ,\label{TreeEqn1}\\
G_{2,full}^{tree}  - G_{3,full}^{tree} \hspace{3mm} &= \hspace{3mm}G^{tree}_{2,eff}\left(a_i^{tree}\right) - G^{tree}_{3,eff}\left(a_i^{tree}\right) \hspace{3mm}&= \ldots + i p^2 a_T^{tree}  + \ldots , \label{TreeEqn2}\\
G_{3,full}^{tree}  \hspace{3mm} &= \hspace{3mm} G^{tree}_{3,eff}\left(a_i^{tree}\right) \hspace{3mm} &= \ldots - i p^2 a_1^{tree}  + \ldots . \label{TreeEqn3}
\end{alignat}
The first equality in Eqs.~(\ref{TreeEqn1})-(\ref{TreeEqn3}) is the matching condition,  while the second equality, which follows from calculating matrix elements of $\{O_i\}$, relates the three different amplitudes $\{G_{i,eff}\}$ to the coefficients of the three different operators $\{O_i\}$ in the effective theory. The ellipses on the RHS denote any non-quadratic dependence on the external momentum $p$, which correspond to operators with dimensions other than 6.

We calculate the full theory amplitudes on the LHS of Eqs.~(\ref{TreeEqn1})-(\ref{TreeEqn3}), then extract its quadratic dependence on $p$ to obtain $a_i^{tree}$. The result is: 
\begin{equation}
\mathcal{L}_{eff}^{0,tree} = \mathcal{L}_{SM} - \frac{2\kappa^2}{M^4}\left(O_T + \frac{1}{2}O_1 - \frac{1}{2}O_2\right) + \ldots ,
\label{L0efftree}
\end{equation}
\begin{equation}
\mathcal{L}_{eff}^{\pm 1, tree} = \mathcal{L}_{SM} + \frac{\kappa^2}{M^4}\left(O_T + O_2\right) + \ldots ,
\label{L1efftree}
\end{equation}
where the ellipses denote higher-dimensional operators and operators that are not relevant for our calculation. Thus,  for the neutral triplet
\begin{equation}
a_{T}^{0, tree}\left(\mu=M\right) = -\frac{2\kappa^2}{M^4}, \hspace{5mm} a_{S}^{0, tree}\left(\mu=M\right) = 0.
\end{equation}
For the charged triplet,
\begin{equation}
a_{T}^{\pm 1, tree}\left(\mu=M\right) = \frac{\kappa^2}{M^4}, \hspace{5mm} a_{S}^{\pm1, tree}\left(\mu=M\right) = 0.
\end{equation}

\subsection*{1-Loop}
Having determined $a_{S,T}^{tree}$, we proceed to calculate $a_{S,T}^{1-loop}$ using Eq.~(\ref{MC1}).  We use the same choices for external momenta and Higgs doublet components as in the tree-level calculation. The 1-loop analogs of Eqs.~(\ref{TreeEqn1})-(\ref{TreeEqn3}) are
\begin{eqnarray}
\left[G_{1full}^{1-loop} - G_{3full}^{1-loop}\right] -   \left[G_{1eff}^{1-loop}(a_i^{tree}) - G_{3eff}^{1-loop}(a_i^{tree})\right] &=& \ldots + i p^2 a_2^{1-loop}  + \ldots, \label{LoopEqn1}\\
\left[G_{2full}^{1-loop} - G_{3full}^{1-loop}\right] -   \left[G_{2eff}^{1-loop}(a_i^{tree}) - G_{3eff}^{1-loop}(a_i^{tree})\right]   &=& \ldots + i p^2 a_T^{1-loop}  + \ldots, \label{LoopEqn2}\\
G_{3full}^{1-loop}- G_{3eff}^{1-loop}(a_i^{tree})  &=& \ldots - i p^2 a_1^{1-loop}  + \ldots . \label{LoopEqn3}
\end{eqnarray}
To obtain $a_T^{1-loop}$, we calculate the amplitudes on the LHS of Eqs.~(\ref{LoopEqn1})-(\ref{LoopEqn3}) and extract the quadratic dependence on external momentum. 
Any non-local contributions in the equations above vanish when the difference between the full and effective theory amplitudes is computed, because these theories have identical behavior in the infrared.  Note that by considering all one-loop diagrams in the full theory, for a given process and external state configuration, we automatically take into account contributions to $a_i^{1-loop}$ that come from all possible wavefunction and vertex renormalizations due to the triplet. We use dimensional regularization and the $\overline{MS}$ prescription in the full and effective theories to regulate UV divergences. All such divergences are cancelled by appropriate counterterms and do not appear in the result for $a_i^{1-loop}$. 

In practice, the $G_{eff}^{1-loop}({a_{i}^{tree})}$ terms in Eqs.~(\ref{LoopEqn1})-(\ref{LoopEqn3}) do not need to be calculated in dimensional regularization further simplifying our approach. This is because we are working in the limit where all SM fields are massless. With massless propagators, the amplitudes $G_{eff}^{1-loop}$ depend on the external momenta only in a non-analytic way. Their only effect in the matching calculation in Eqs.~(\ref{LoopEqn1})-(\ref{LoopEqn3}) is to cancel all non-analytic terms of $G_{full}^{1-loop}$. We thus do not  compute effective theory diagrams.

Extracting the coefficient $a_S^{1-loop}$ is considerably simpler, because $O_S$ is the only CP-conserving dimension-6 operator composed of two Higgs fields, one $SU(2)$ gauge boson, and one $U(1)$ gauge boson.  Let $D_{\mu\nu}\left(p\right)$ denote the amplitude for the scattering process
\begin{equation}
H_1 A_\mu^3 B_\nu \longrightarrow H_1,
\label{Sprocess}
\end{equation}
where both Higgs lines have zero momentum, and $A_\mu^3$ and $B_\mu$ have momenta $p$ and $-p$, respectively. Another straightforward calculation gives
\begin{equation}
\frac{1}{2(d-1)} \left[ \left(D^{\mu}_{\phantom{\mu}\mu}\right)^{1-loop}_{full} \left(p\right) -  \left(D^{\mu}_{\phantom{\mu}\mu}\right)^{1-loop}_{eff} \left(p,a_i^{tree}\right)\right] =\ldots + ip^2 a_S^{1-loop} + \ldots
\label{S1loop}
\end{equation}
where $d$ is the dimension of spacetime.  To obtain $a_S^{1-loop}$, we follow the same steps used for computing $a_T^{1-loop}$: we calculate the 1-loop amplitude on the LHS of Eq.~(\ref{S1loop}) and extract the quadratic term in $p$.

Carrying out these steps, we get the 1-loop corrections to $a_{S,T}$: 
\begin{eqnarray}
a_{T}^{0, 1-loop}(\mu=M) &=& \frac{1}{(4\pi)^{2}}\frac{\kappa^{2}}{M^{4}}\left(-\frac{3}{2}\lambda+16\eta-\frac{37}{4}g_{2}^{2}+5\frac{\kappa^{2}}{M^{2}}\right) \label{AT1n}, \\
a_{T}^{\pm 1, 1-loop}(\mu=M) &=& \frac{1}{(4\pi)^{2}}\frac{\kappa^{2}}{M^{4}}\left(\frac{3}{4}\lambda+\frac{11}{8}g_{1}^{2}+\frac{37}{8}g_{2}^{2}-\frac{17}{3}\frac{\kappa^{2}}{M^{2}}-\frac{22}{3}\eta_{2}-4\eta_{1}\right) \nonumber \\ &&  - \frac{2}{3}\frac{1}{(4\pi)^2}\frac{\eta_2^2}{M^2} \label{AT1c},\\
a_{S}^{0, 1-loop}(\mu=M) &=& -\frac{1}{(4\pi)^{2}}\frac{g_{1}g_{2}}{M^{2}}\left(\frac{1}{120}g_{2}^{2}+\frac{5}{24}\frac{\kappa^{2}}{M^{2}}\right) \label{AS1n},\\
a_{S}^{\pm1, 1-loop}(\mu=M) &=& \frac{1}{(4\pi)^{2}}\frac{g_{1}g_{2}}{M^{2}}\left(\frac{1}{3}\eta_{2}-\frac{1}{40}g_{1}^{2}-\frac{1}{60}g_{2}^{2}+\frac{1}{8}\frac{\kappa^{2}}{M^{2}}\right)\label{AS1c}.
\end{eqnarray}

\section{Running}
\label{app:running}

In Appendix~\ref{app:matching}, we described the matching procedure for determining the EFT coefficients $a_{S,T}(\mu=M)$. In this appendix, we briefly review the procedure for calculating the RG running of these coefficients down to $v$. Since we are interested in one-loop accuracy, only the running of the tree-level part of $a_i(\mu=M)$ is needed. To leading order in $\mathrm{log}\left(\frac{v}{M}\right)$, the final answer for $a_i$ takes the form
\begin{equation}
a_i(\mu = v) = a_i^{tree}\left(\mu=M\right) + a_i^{1-loop}\left(\mu=M\right) + \beta_i \,\mathrm{log}\left(\frac{v}{M}\right),
\label{AGNL}
\end{equation}
where $\beta_i$ is the 1-loop beta function.

Under the RG running, different dimension-6 operators mix, so operators that did not appear at the matching scale  can be radiatively generated from the ones that are present there. As we did previously, radiative corrections to $O_{S,T}$ can be extracted using the methods described in Appendix~\ref{app:matching}. Let the superscript $RG$ denote the UV divergent part in the $\overline{MS}$ scheme in dimensional regularization of a 1-loop vertex renormalization diagram in the effective theory. Then, again using the notation $G_{1,2,3}$ from Eq.~(\ref{G}) and $D_{\mu\nu}$ defined above Eq.~(\ref{Sprocess}), we have
\begin{equation}
G_{2}^{RG}-G_{3}^{RG} =\ldots  -i p^2 a_{T}(Z_{T}Z_{H}^{2}-1) + \ldots ,
\label{TRG}
\end{equation}
\begin{equation}
\frac{1}{2(d-1)} \left(D^{RG}\right)^{\mu}_{\phantom{\mu}\mu} =\ldots  -i p^2 a_{S}(Z_{S}Z_{H}Z_{A}^{1/2}Z_{B}^{1/2}-1) + \ldots .
\label{SRG}
\end{equation}
Here, $Z_{H,A,B}$ are the $Z$-factors for the wavefunction renormalization of $H$, $A_\mu^a$, and $B_\mu$, which are straightforward to calculate, while $Z_{S,T}$ are the Z-factors associated with renormalization of $O_{S,T}$ and are defined by Eqs.~(\ref{TRG})-(\ref{SRG}). These equations are just the statement that $Z_{S,T}$ cancel the divergences of 1-loop diagrams that renormalize $O_{S,T}$. As before, the ellipses denote non-quadratic powers of $p$.  

The beta functions, $\beta_{S,T}$, for $a_{S,T}$ are related to the Z-factors by
\begin{equation}
\beta_\xi = -a_\xi \frac{1}{Z_\xi} \frac{\mathrm{d}}{\mathrm{d}\,\mathrm{log}\mu} Z_\xi, \hspace{5mm} \xi=S,T.
\label{beta}
\end{equation}
Calculating $Z_{S,T}$ using Eqs.~(\ref{TRG})-(\ref{SRG}), we find the following beta functions for the neutral and charged triplet cases:
\begin{eqnarray}
\beta_{T}^{0} & = & - \frac{2}{(4\pi)^{2}}\left(3\lambda-3g_{1}^{2}+\frac{9}{2}g_{2}^{2}+24y_{B}^{2}+24y_{T}^{2}\right)\frac{\kappa^2}{M^4}, \\
\beta_{T}^{\pm1} & = & \frac{1}{(4\pi)^{2}}\left(3\lambda+\frac{3}{2}g_{1}^{2}+\frac{9}{2}g_{2}^{2}+24y_{B}^{2}+24y_{T}^{2}\right)\frac{\kappa^2}{M^4}, \\
\beta_{S}^{0} & = & -  \frac{1}{6}\frac{g_1 g_2}{(4\pi)^{2}} \frac{\kappa^2}{M^4}, \\
\beta_{S}^{\pm1} & = & \frac{1}{3}\frac{g_1 g_2}{(4\pi)^{2}} \frac{\kappa^2}{M^4}.
\end{eqnarray}
Note that $\beta_S \propto a_T$, as a consequence of operator mixing. Combining these results with the results of matching gives the final answers in Eqs.~(\ref{FRSTN})-(\ref{FRSSC}). Note that the expressions in the neutral and charged cases are different because the tree-level matching coefficients of the operators $O_i$, $i=1,2,T$,
differ in these two cases.

\section{Explicit Examples}
\label{app:sample}

\subsection*{Example of matching: $\eta_2^2$ contribution to $a_T^{\pm1}$.}
In this example, we consider the case of the charged triplet and calculate the contribution to $a_T^{\pm1}$ proportional to $\eta_2^2$ in Eq.~(\ref{FRSTC}). This contribution is important, because it is the only $\kappa$-independent contribution to $a_T^{\pm 1}$, the implications of which are discussed in Section~\ref{sec:results}.

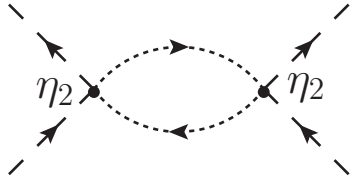
\begin{figure}[hp]
\fcolorbox{white}{white}{
  \begin{picture}(142,66) (239,-159)
    \SetWidth{1.0}
    \SetColor{Black}
    \Line[dash,dashsize=10,arrow,arrowpos=0.5,arrowlength=5,arrowwidth=2,arrowinset=0.2](240,-158)(272,-126)
    \Line[dash,dashsize=10,arrow,arrowpos=0.5,arrowlength=5,arrowwidth=2,arrowinset=0.2](272,-126)(240,-94)
    \Arc[dash,dashsize=2,arrow,arrowpos=0.5,arrowlength=5,arrowwidth=2,arrowinset=0.2,clock](304,-150)(40,143.13,36.87)
    \Arc[dash,dashsize=2,arrow,arrowpos=0.5,arrowlength=5,arrowwidth=2,arrowinset=0.2,clock](304,-102)(40,-36.87,-143.13)
    \Line[dash,dashsize=10,arrow,arrowpos=0.5,arrowlength=5,arrowwidth=2,arrowinset=0.2](368,-158)(336,-126)
    \Line[dash,dashsize=10,arrow,arrowpos=0.5,arrowlength=5,arrowwidth=2,arrowinset=0.2](336,-126)(368,-94)
    \Vertex(272,-127){2.236}
    \Vertex(336,-127){2.236}
    \Text(251,-133)[lb]{\Large{\Black{$\eta_2$}}}
    \Text(346,-131)[lb]{\Large{\Black{$\eta_2$}}}
  \end{picture}
}
\caption{The 1-loop process giving rise to the $\kappa$-independent term in $a^{\pm 1}_T$. Long dashed lines represent the  Higgs doublets, while the short dashed lines  represent the heavy triplet.}
\label{ETAG}
\end{figure}

The full-theory topology giving rise to the $\eta_2^2$ contribution is shown in Fig.~\ref{ETAG}. Labeling the momenta and components of the external Higgses in the same way as in Fig.~\ref{2Higgs}, and noting that there are two possible permutations of the external lines in Fig.~\ref{ETAG}, the integral expression for the diagram is
\begin{eqnarray}
G_{s_{1},s_{2},s_{3},s_{4}}(p_{1},p_{2},p_{3},p_{4})& = &2\left(2\delta_{s_{1}s_{4}}\delta_{s_{2}s_{3}}-\delta_{s_{1}s_{3}}\delta_{s_{2}s_{4}}\right)\eta_{2}^{2}\int\frac{d^{d}\ell}{(2\pi)^{d}}\frac{1}{\ell^{2}-M^{2}}\frac{1}{(\ell+p_{1}-p_{3})^{2}-M^{2}} \nonumber \\ && + \left(p_3,s_3 \leftrightarrow p_4,s_4 \right),
\end{eqnarray}
where $d=4-2\epsilon$ is the dimension of spacetime.
 
With this expression in hand, we can now use Eqs.~(\ref{G}) and (\ref{LoopEqn2}) to solve for the contribution to $a_T^{\pm 1}$. This requires extracting the $p^2$ term on the LHS of Eq.~(\ref{LoopEqn2}). A useful intermediate result for expanding loop integrands in powers of $p^2$ is 
\begin{eqnarray}
\hspace{-20pt} \frac{1}{(\ell+p)^2-M^2} &= & \hspace{-5pt}\frac{1}{\ell^2-M^2} + \frac{dM^2+(4-d)\ell^2}{d(\ell^2-M^2)^3}p^2 \nonumber \\ 
&&\hspace{-5pt} + \frac{d(d+2)M^4+2(6-d)(d+2)M^2\ell^2+(6-d)(4-d)(\ell^2)^2}{d(d+2)(\ell^2-M^2)^5}(p^2)^2+\ldots .
\end{eqnarray}
Once an integrand is expanded in powers of $p^2$, all loop integrals are easily evaluated via Feynman parameters.

For the diagram in Fig.~\ref{ETAG}, Eq.~(\ref{LoopEqn2}) gives
\begin{equation}
\delta \left[i p^2 a_T^{\pm 1, 1-loop} \right]_{\mathrm{Fig.}~\ref{ETAG}} = \left[ G_2 - G_3 \right]_{p^{2}\, part}=-\frac{2}{3}\eta_{2}^{2}\frac{ip^{2}}{(4\pi)^{2}M^{2}},
\end{equation}
where we have used dimensional regularization in the $\overline{MS}$ scheme. Consequently, 
\begin{equation}
\delta \left[a_T^{\pm1, 1-loop} \right]_{\mathrm{Fig.}~\ref{ETAG}} = -\frac{2}{3}\frac{\eta_{2}^{2}}{(4\pi)^{2}M^{2}}.
\end{equation}
This corresponds to the last term in Eq.~(\ref{FRSTC}) and makes a positive contribution to $T$, as discussed in Section~\ref{sec:results}.

\subsection*{Example of running: RG-running of $a_S^{\pm 1}$}

In this example, we consider the case of the charged triplet and calculate the RG-running of $a_S^{\pm1}$. We compute the beta function, $\beta_S$, appearing in Eq.~(\ref{AGNL}) for $a_S^{\pm1}$. This example illustrates the procedure for RG-running and for extracting contributions to the $S$ parameter. 

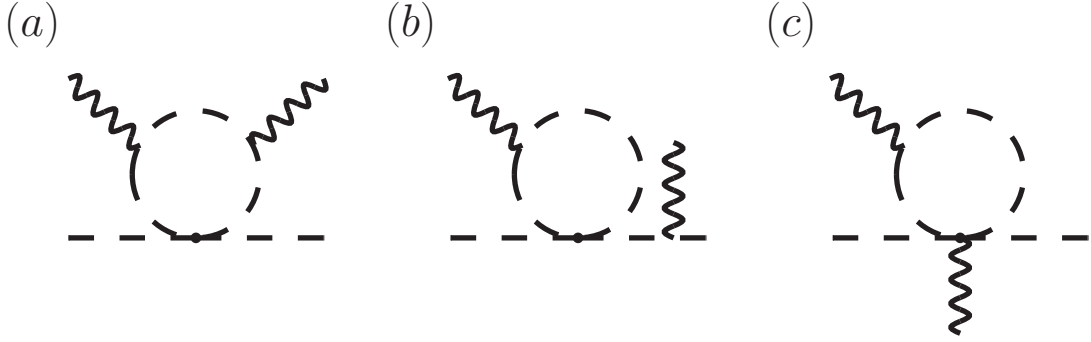
\begin{figure}[h]
\fcolorbox{white}{white}{
  \begin{picture}(406,131) (99,-63)
    \SetWidth{2.0}
    \SetColor{Black}
    \Arc[dash,dashsize=10](168,-1)(24,180,540)
    \Line[dash,dashsize=10](216,-25)(168,-25)
    \Line[dash,dashsize=10](120,-25)(168,-25)
    \Photon(147,10)(120,35){3.5}{4}
    \Photon(190,10)(216,35){3.5}{4}
    \Text(96,47)[lb]{\Large{\Black{$(a)$}}}
    \Text(240,47)[lb]{\Large{\Black{$(b)$}}}
    \Text(384,47)[lb]{\Large{\Black{$(c)$}}}
    \Arc[dash,dashsize=10](312,-1)(24,180,540)
    \Line[dash,dashsize=10](264,-25)(312,-25)
    \Line[dash,dashsize=10](360,-25)(312,-25)
    \Photon(291,10)(263,35){3.5}{4}
    \Photon(348,-25)(348,11){3.5}{4}
    \Arc[dash,dashsize=10](456,-1)(24,180,540)
    \Photon(435,10)(407,35){3.5}{4}
    \Photon(456,-25)(456,-61){3.5}{4}
    \Line[dash,dashsize=10](408,-25)(456,-25)
    \Line[dash,dashsize=10](504,-25)(456,-25)
    \SetWidth{1.0}
    \Vertex(168,-25){2}
    \Vertex(312,-25){2}
    \Vertex(456,-25){2}
  \end{picture}
}

\caption{All effective-theory 1-loop topologies contributing to the renormalization of $O_S$. Dashed lines represent the Higgs doublet.}
\label{AppCSRG}
\end{figure}

Recall that after integrating out the charged triplet at tree-level, we are left with the effective Lagrangian in Eq.~(\ref{L1efftree}). Thus, the Feynman rules in the effective theory are those of the SM plus new vertices due to the tree-level presence of $O_T$ and $O_2$. These additional vertices are comprised of four Higgses and either zero, one, or two gauge bosons. For our example, we will need the new four-Higgs vertex, which we call $V_{s_{1},s_{2},s_{3},s_{4}}(p_{1},p_{2},p_{3},p_{4})$, where $\left\{p_j,s_j\right\}$, $j=1,2$, denote the incoming Higgs momenta and its components, while $\left\{p_j,s_j\right\}$, $j=3,4$, denote the outgoing ones, in analogy with Fig.~\ref{2Higgs}. The amplitude for this vertex is
\begin{equation}
V_{s_{1},s_{2},s_{3},s_{4}}(p_{1},p_{2},p_{3},p_{4}) = \left(\delta_{s_1s_3}\delta_{s_2s_4} + \delta_{s_1s_4}\delta_{s_2s_3}\right) \frac{i\kappa^2}{M^4}\left(p_1 + p_2\right)^2.
\end{equation}
Although $O_S$ does not appear at tree-level in Eq.~(\ref{AGNL}), the new effective vertices generate $O_S$ in RG-running. In particular, Fig.~\ref{AppCSRG} shows the 1-loop topologies that contribute to the process $D_{\mu\nu}(p)$ (Eq.~(\ref{Sprocess})) and thus correct $a_S^{\pm1,tree}=0$. Note the 4-Higgs and 4-Higgs-1-gauge-boson vertices in these diagrams.

We consider the contribution of Fig.~\ref{AppCSRG}(a). In accordance with Eq.~(\ref{Sprocess}), the amplitude involves upper components of external Higgses with  zero momentum and external gauge bosons $A_\mu^3$, $B_\nu$ with momenta $\pm p$. There are two ways of attaching the gauge bosons.  Summing both possibilities gives the following contribution to $D_{\mu\nu}(p)$:
\begin{eqnarray}
\delta \left[D_{\mu\nu}(p)\right]_{\mathrm{Fig.}~\ref{AppCSRG}(a)}& = &\frac{ig_1g_2}{4}\int \frac{\mathrm{d}^d\ell}{(2\pi)^d} V_{1,s,s',1}(0,\ell,\ell,0)\,\sigma^3_{s,s'} \frac{1}{(\ell^2)^2}\frac{1}{(\ell+p)^2}\left(2\ell+p\right)_\mu \left(2\ell+p\right)_\nu \nonumber \\ && + \left(p \rightarrow -p\right).
\label{Eg1}
\end{eqnarray}  
We now  contract Lorentz indices and expand in $p$ to find the $p^2$ term. We only need the UV divergent part for the $\beta$ function:
\begin{align}
\delta \left[\frac{1}{2(d-1)}D^\mu_{\phantom{\mu}\mu}(p)\right]_{\mathrm{Fig.}~\ref{AppCSRG}(a)} &= \left(ip^2\right) \frac{ig_1g_2\kappa^2}{M^4}\frac{(2-\frac{3}{4}d)}{d(d-1)} \int \frac{\mathrm{d}^d\ell}{(2\pi)^d} \frac{1}{(\ell^2)^2} + \ldots \notag \\
&\stackrel{UV}{\longrightarrow} \left(ip^2\right) \left(\frac{1}{12} \frac{g_1g_2\kappa^2}{(4\pi)^2 M^4} \frac{1}{\bar{\epsilon}}\right) + \ldots,  \label{Eg3}
\end{align}
where $\frac{1}{\bar{\epsilon}} = \frac{1}{\epsilon} - \gamma + \mathrm{log}4\pi$, and the ellipses denote non-quadratic powers of $p$. 
 
In a similar manner, one needs to find the contributions from the remaining two topologies in Fig.~\ref{AppCSRG}. We simply state the result:
\begin{align}
\delta \left[\frac{1}{2(d-1)}(D^{RG})^\mu_{\phantom{\mu}\mu} \right]_{\mathrm{Fig.}~\ref{AppCSRG}(b)} &= 0 +\ldots , \label{Topb}\\
\delta \left[\frac{1}{2(d-1)}(D^{RG})^\mu_{\phantom{\mu}\mu} \right]_{\mathrm{Fig.}~\ref{AppCSRG}(c)} &= -ip^2 \frac{1}{4} \frac{g_1g_2\kappa^2}{(4\pi)^2 M^4} \frac{1}{\bar{\epsilon}} +\ldots . \label{Topc}
\end{align}
Summing Eqs.~(\ref{Topa})-(\ref{Topc}) gives the full contribution to the LHS of Eq.~(\ref{SRG}). On the RHS, $Z_{H,A,B}$ are the standard wavefunction renormalization $Z$-factors, which in our conventions are given by
\begin{align}
Z_H &= 1 + \frac{1}{(4\pi)^2}\left[\frac{1}{2}g_1^2 + \frac{3}{2}g_2^2 - 6\left(y_T^2 + y_B^2\right)\right]\frac{1}{\bar{\epsilon}}, \\
Z_A &= 1 - \frac{29}{6} \frac{g_2^2}{(4\pi)^2} \frac{1}{\bar{\epsilon}}, \\
Z_B &= 1 - \frac{27}{2} \frac{g_1^2}{(4\pi)^2} \frac{1}{\bar{\epsilon}}.
\end{align}
In this example, since $a_S^{\pm1,tree}=0$, it suffices to take $Z_{H,A,B}=1$, but we stated the full 1-loop answers for completeness.
Now, using Eqs.~(\ref{SRG})-(\ref{beta}) one can solve for $Z_S$ and $\beta_S$, respectively, to obtain
\begin{equation}
\beta_S = \frac{1}{3}\frac{g_1g_2}{(4\pi)^2} \frac{\kappa^2}{M^4},
\end{equation}
which corresponds to the last term in Eq.~(\ref{FRSSC}).


\end{document}